\documentclass[aps,pra,twocolumn,showpacs,superscriptaddress]{revtex4}
\usepackage{amsfonts}
\usepackage{amsmath}
\usepackage{amssymb}
\usepackage{graphicx}

\begin{document}

\title{Structure and stability of quasi-two-dimensional boson-fermion
mixtures with vortex-antivortex superposed states}
\author{Linghua Wen}
\email{wenlinghua@lcu.edu.cn}
\affiliation{Institute of Physics,
Chinese Academy of Sciences, Beijing 100190, China}
\affiliation{Department of Physics, Liaocheng University, Shandong
252059, China}
\author{Yongping Zhang}
\affiliation{Institute of Physics, Chinese Academy of Sciences,
Beijing 100190, China}
\affiliation{Department of Physics and
Astronomy, Washington State University, Pullman, Washington, 99164
USA}
\author{Jian Feng}
\affiliation{Department of Physics, Liaocheng University, Shandong 252059, China}
\date{\today }

\begin{abstract}
We investigate the equilibrium properties of a quasi-two-dimensional
degenerate boson-fermion mixture (DBFM) with a bosonic vortex-antivortex
superposed state (VAVSS) using a quantum-hydrodynamic model. We show that,
depending on the choice of parameters, the DBFM with a VAVSS can exhibit
rich phase structures. For repulsive boson-fermion (BF) interaction, the
Bose-Einstein condensate (BEC) may constitute a petal-shaped
\textquotedblleft core\textquotedblright\ inside the honeycomb-like
fermionic component, or a ring-shaped joint \textquotedblleft
shell\textquotedblright\ around the onion-like fermionic cloud, or multiple
segregated \textquotedblleft islands\textquotedblright\ embedded in the
disc-shaped Fermi gas. For attractive BF interaction just below the
threshold for collapse, an almost complete mixing between the bosonic and
fermionic components is formed, where the fermionic component tends to mimic
a bosonic VAVSS. The influence of an anharmonic trap on the density
distributions of the DBFM with a bosonic VAVSS is discussed. In addition, a
stability region for different cases of DBFM (without vortex, with a bosonic
vortex, and with a bosonic VAVSS) with specific parameters is given.
\end{abstract}

\maketitle

\section{Introduction}

With the experimental realizations of atomic Bose-Einstein condensate (BEC),
molecular BEC and fermionic condensate, considerable attention has been
given to degenerate boson-fermion mixtures (DBFMs) where particles obeying
different quantum statistics are intermingled \cite{Giorgini,Pethick}. The
DBFMs, such as $^{7}$Li-$^{6}$Li \cite{Truscott,Schreck}, $^{23}$Na-$^{6}$Li
\cite{Hadzibabic}, $^{87}$Rb-$^{40}$K \cite%
{Modugno,Roati,Goldwin,Ospelkaus,Best}, and $^{174}$Yb-$^{173}$Yb \cite%
{Fukuhara}, have been successfully observed by different experimental groups
via interspecies sympathetic cooling. In addition, such mixtures can also be
realized from an imbalanced two-component fermi gas where all minority
fermions pair up with majority fermions and form a BEC \cite%
{Partridge,Zwierlein}.

The collisional interactions between bosons and fermions strongly affect the
properties of DBFMs, even though the ultracold mixed gases are very dilute
\cite%
{Molmer,Roth1,Adhikari1,Akdeniz,Minguzzi,Capuzzi,Nygaard,Viverit,Roth2,Jezek,Adhikari2,Ma,Hu,Santhanam,Salasnich1,Fang,Maruyama,Marchetti}%
. A noticeable example is the structure and stability of a trapped DBFM
which has been investigated extensively by several authors. In the case of
repulsive boson-fermion (BF) interaction, the system is expected to undergo
a mixing-demixing transition with the increase of the repulsion, while in
the case of attractive BF interaction the mixture collapses above a critical
strength \cite%
{Modugno,Ospelkaus,Molmer,Roth1,Adhikari1,Akdeniz,Minguzzi,Capuzzi,ModugnoM,Chui,Takeuchi,Linder}%
. In particular, the effect of a bosonic vortex on the stability and the
mixing-demixing transition of a DBFM is recently addressed \cite%
{Jezek,Adhikari3}. It is shown that when the BEC sustains a vortex state the
DBFM becomes more stable.

In this paper, we consider the equilibrium properties of a
quasi-two-dimensional (quasi-2D) DBFM in the presence of a bosonic
vortex-antivortex superposed state (VAVSS). Recently, the VAVSS in BECs has
attracted extensive research interest as it exhibits peculiar petal-like
structure and rich dynamics \cite{Kapale,Liu,Simula,Thanvanthri,Wen1}.
Furthermore, the VAVSS may have fundamental as well as practical
applications in quantum information and inertial sensing. Experimentally,
the creation of VAVSS in BECs has been reported by Anderson \textit{et al}
\cite{Andersen} and Wright \textit{et al }\cite{Wright}. Throughout the
present work, we use a quantum-hydrodynamic model \cite{Adhikari3} to
investigate the structure of the exact 2D spatial density distributions and
the stability of the system against collapse. This mean-field-hydrodynamic
model is quite successful in the study of collapse dynamics \cite{Adhikari1}
and dark \cite{Adhikari4} and bright solitons \cite{Adhikari2,Salasnich2} in
a DBFM. The theoretical predictions on fermionic collapse in a DBFM agree
well with the experimental results \cite{Modugno,Ospelkaus}, and those on
bright solitons are in agreement with a relevant microscopic investigation
\cite{Karpiuk}. We show that, for interspecies repulsion, the bosonic
component may form a petal-shaped \textquotedblleft core\textquotedblright\
inside the honeycomb-like fermionic one, or a ring-shaped joint
\textquotedblleft shell\textquotedblright\ around the onion-like fermionic
cloud, or even segregated \textquotedblleft islands\textquotedblright\
embedded in the disc-shaped Fermi gas. For interspecies attraction below the
critical value for collapse, there is an enhanced mixing (even an almost
complete mixing) between bosons and fermions, where the fermionic cloud
tends to simulate a bosonic VAVSS. Moreover, we find that the DBFM with a
bosonic VAVSS and the one with a bosonic pure vortex sustain an intermittent
stability diagram.

The paper is organized as follows. In section II we introduce the
mean-field-hydrodynamic model for a quasi-2D DBFM, which is comprised of two
coupled quantum-hydrodynamic equations. In section III, we present the
numerical results and some discussion on the structure of the exact 2D
density profiles of a DBFM with a bosonic VAVSS. Furthermore, a stability
region for different cases of DBFM with specific parameters is displayed.
The conclusion is outlined in the last section.

\section{Mean-field-hydrodynamic model for a degenerate boson-fermion mixture%
}

We consider a dilute degenerate mixture composed of $N_{B}$ condensed bosons
and $N_{F}$ spin-polarized fermions at zero temperature, with respective
particle mass $m_{B}$ and $m_{F}$. In the case of strong axial ($z$%
-direction) confinement, the DBFM reduces to a quasi-2D system, where the
axial trap frequency is denoted by $\omega _{z}$. Here we use the
time-dependent mean-field-hydrodynamic model developed in Ref. \cite%
{Adhikari3} to describe this BF system. The Lagrangian density of the DBFM
reads%
\begin{equation}
\mathcal{L}=\mathcal{L}_{\mathit{B}}+\mathcal{L}_{\mathit{F}}+\mathcal{L}_{%
\mathit{BF}}.  \label{Lagrangian density}
\end{equation}%
Here $\mathcal{L}_{\mathit{B}}$ is the ordinary bosonic Lagrangian density,
\begin{eqnarray}
\mathcal{L}_{\mathit{B}} &=&\frac{i\hbar }{2}(\Psi _{B}^{\ast }\frac{%
\partial \Psi _{B}}{\partial t}-\Psi _{B}\frac{\partial \Psi _{B}^{\ast }}{%
\partial t})-\frac{\hbar ^{2}}{2m_{B}}(\left\vert \frac{\partial \Psi _{B}}{%
\partial x}\right\vert ^{2}  \notag \\
&&+\left\vert \frac{\partial \Psi _{B}}{\partial y}\right\vert
^{2})-V_{B}n_{B}-\frac{g_{BB}}{2}n_{B}^{2},  \label{bosonic Lagrangian}
\end{eqnarray}%
where $\Psi _{B}(x,y,t)$ is the hydrodynamic field of the Bose gas, i.e.,
the macroscopic BEC wave function, $n_{B}=\left\vert \Psi _{B}\right\vert
^{2}$ is the 2D bosonic density with normalization $N_{B}=\iint \left\vert
\Psi _{B}\right\vert ^{2}dxdy$, and $V_{B}$ is the external trapping
potential for the bosons. $g_{BB}=2\sqrt{2\pi }\hbar
^{2}a_{BB}/(a_{zB}m_{B}) $ is the 2D interatomic interaction strength with $%
a_{BB}$ being the 3D boson-boson (BB) $s$-wave scattering length and $a_{zB}=%
\sqrt{\hbar /(m_{B}\omega _{z})}$ the axial harmonic length of bosons. The
bosonic Lagrangian density $\mathcal{L}_{\mathit{B}}$ can describe well all
the dynamical properties of the dilute quasi-2D BEC (see Ref. \cite%
{Adhikari3} and references therein). For simplicity, here we have assumed
the BEC to be strictly 2D.

The Lagrangian density of the Fermi gas
$\mathcal{L}_{\mathit{F}\text{ }}$is expressed by \cite{Adhikari3}
\begin{eqnarray}
\mathcal{L}_{\mathit{F}} &=&\frac{i\hbar }{2}(\Psi _{F}^{\ast }\frac{%
\partial \Psi _{F}}{\partial t}-\Psi _{F}\frac{\partial \Psi _{F}^{\ast }}{%
\partial t})-\frac{\hbar ^{2}}{6m_{F}}(\left\vert \frac{\partial \Psi _{F}}{%
\partial x}\right\vert ^{2}  \notag \\
&&+\left\vert \frac{\partial \Psi _{F}}{\partial y}\right\vert
^{2})-V_{F}n_{F}-\xi _{F}(n_{F}),  \label{fermionic Lagrangian}
\end{eqnarray}%
where $\Psi _{F}(x,y,t)$ is the hydrodynamic field of the degenerate Fermi
gas (DFG) (i.e., an average fermionic wave function), $n_{F}=\left\vert \Psi
_{F}\right\vert ^{2}$ is the 2D fermionic density with normalization $%
N_{F}=\iint \left\vert \Psi _{F}\right\vert ^{2}dxdy$, and $V_{F}$ is the
external trapping potential for\ the fermions. $\xi _{F}(n_{F})$ denotes the
zero-temperature bulk energy density of an ideal quasi-2D Fermi gas under
axial harmonic confinement, and it reads%
\begin{equation}
\xi _{F}=\left\{
\begin{array}{ll}
\hbar \omega _{z}\pi (n_{F}a_{zF})^{2},\quad \quad 0\leqslant
n_{F}a_{zF}^{2}<{\frac{1}{2\pi },} &  \\
\frac{\hbar \omega _{z}}{6\pi a_{zF}^{2}}(4\pi n_{F}a_{zF}^{2}-1)^{3/2}+{%
\frac{1}{12\pi },}\quad \quad n_{F}a_{zF}^{2}\geqslant {\frac{1}{2\pi }}\,.
&
\end{array}%
\right.  \label{energy density of Fermi gas}
\end{equation}%
For $0\leqslant n_{F}<{1/(2\pi a_{zF}^{2})}$, the Fermi gas is strictly 2D,
where $a_{zF}=\sqrt{\hbar /(m_{F}\omega _{z})}$ is the axial harmonic length
of fermions. For $n_{F}\geqslant {1/(2\pi a_{zF}^{2})}$, the Fermi gas has
the 2D-3D crossover, where several single-particle modes of the harmonic
oscillator along the $z$ axis are occupied. The Fermi gas becomes 3D for $%
n_{F}\gg {1/(2\pi a_{zF}^{2})}$. Due to the Pauli exclusion principle, the
interaction between identical fermions in spin polarized state is highly
suppressed and has been neglected in the Lagrangian density $\mathcal{L}_{%
\mathit{F}}$ and will be neglected throughout this paper. The second term $-$
$\hbar ^{2}\left\vert \nabla _{x,y}\Psi _{F}\right\vert ^{2}/6m_{F}$ in Eq.(%
\ref{fermionic Lagrangian}) represents the Thomas-Fermi-Weizs\"{a}cker
kinetic energy which includes the surface kinetic energy due to spatial
variation \cite{Capuzzi}. Note that the fermionic Lagrangian density $%
\mathcal{L}_{\mathit{F}\text{ }}$ can only be safely used to describe the
static and collective properties of the quasi-2D Fermi gas \cite{Adhikari3}.
As the strict 2D Fermi gas requiring a very small number of particles is not
the case of real experiments, we take into account the 2D-3D crossover in
the fermionic component.

It may not be entirely proper to define an average fermionic wave
function $\Psi _{F}$ for a DFG. A proper treatment of the DFG should
be performed using a fully antisymmetrized many-body Slater
determinant wave function. However, the probability density $n_{F}$
of a DFG calculated in this way should lead to reasonable results
\cite{Molmer} and has led to proper probability distribution for a
DFG \cite{Adhikari1,Capuzzi} as well as results for collapse of a
DFG \cite{Adhikari1,Capuzzi,ModugnoM} in agreement with experiment.
This method has also been used successfully to
predict fermionic bright and dark solitons in a DBFM \cite%
{Adhikari2,Adhikari4}. The virtue of the mean-field-hydrodynamic model for a
DFG over a microscopic description is its simplicity and good predictive
power. Similar mean-field treatment for degenerate fermions can also be
found in Refs. \cite{Andreev,Subasi}.

Finally, the Lagrangian density $\mathcal{L}_{\mathit{BF}}$ for BF
interaction is given by%
\begin{equation}
\mathcal{L}_{\mathit{BF}}=-g_{BF}n_{B}n_{F},
\label{boson-fermion interaction}
\end{equation}%
where $g_{BF}=2\pi a_{BF}\hbar ^{2}/(m_{BF}\sqrt{2\pi a_{zB}a_{zF}})$ is the
coupling constant of BF interaction in terms of the 3D BF $s$-wave
scattering length $a_{BF}$ and the BF reduced mass $%
m_{BF}=m_{B}m_{F}/(m_{B}+m_{F})$.

In terms of Eqs. (\ref{Lagrangian density})-(\ref{boson-fermion interaction}%
), the Euler-Lagrange equations of motion become%
\begin{eqnarray}
i\hbar \frac{\partial }{\partial t}\Psi _{B} &=&[-\frac{\hbar ^{2}}{2m_{B}}(%
\frac{\partial ^{2}}{\partial x^{2}}+\frac{\partial ^{2}}{\partial y^{2}}%
)+V_{B}+g_{BB}\left\vert \Psi _{B}\right\vert ^{2}  \notag \\
&&+g_{BF}\left\vert \Psi _{F}\right\vert ^{2}]\Psi _{B},
\label{boson equation} \\
i\hbar \frac{\partial }{\partial t}\Psi _{F} &=&[-\frac{\hbar ^{2}}{6m_{F}}(%
\frac{\partial ^{2}}{\partial x^{2}}+\frac{\partial ^{2}}{\partial y^{2}}%
)+V_{F}+\mu _{F}  \notag \\
&&+g_{BF}\left\vert \Psi _{B}\right\vert ^{2}]\Psi _{F}.
\label{fermion equation}
\end{eqnarray}%
Here $\mu _{F}=\partial \xi _{F}/\partial n_{F}$ is the bulk chemical
potential of the non-interacting Fermi gas in the 2D-3D crossover, which is
expressed by%
\begin{equation}
\mu _{F}=\left\{
\begin{array}{ll}
2\pi \hbar \omega _{z}n_{F}a_{zF}^{2},\quad \quad 0\leqslant n_{F}a_{zF}^{2}<%
{\frac{1}{2\pi },} &  \\
\hbar \omega _{z}\sqrt{4\pi n_{F}a_{zF}^{2}-1}{,}\quad \quad
n_{F}a_{zF}^{2}\geqslant {\frac{1}{2\pi }}\,. &
\end{array}%
\right.  \label{chemical potential of Fermi gas}
\end{equation}%
The resultant quantum-hydrodynamic equations have a nonpolynomial
nonlinearity for the fermionic component as shown in Eqs. (\ref{boson
equation}), (\ref{fermion equation}) and (\ref{chemical potential of Fermi
gas}).

For simplicity, we assume that the 2D radial trapping potentials for the
bosonic and fermionic components are the same and take the forms%
\begin{equation}
V_{B}(x,y)=V_{F}(x,y)=\frac{1}{2}m_{B}\omega _{\bot }^{2}\left[
x^{2}+y^{2}+\lambda \frac{(x^{2}+y^{2})^{2}}{a_{0}^{2}}\right] ,
\label{trapping potential}
\end{equation}%
where $a_{0}=$ $\sqrt{\hbar /(m_{B}\omega _{\bot })}$ is the radial harmonic
length of bosons with $\omega _{\bot }$ the radial trap frequency as seen by
the bosons. The quartic term in Eq.(\ref{trapping potential}) denotes the
anharmonicity of the traps, which is possibly caused by experimental
uncertainties or artificial quartic confinement. $\lambda $ is a
dimensionless parameter that determines the relative strength of the quartic
component for bosons. Our choice of the trapping potentials corresponds to a
reduction of $\omega _{\bot }$ by a factor $\sqrt{m_{B}/m_{F}}$ as in the
study by Modugno \textit{et al} \cite{ModugnoM} and a reduction of $\lambda $
by a factor $\sqrt{m_{B}/m_{F}}$, where the two assumptions give a simpler
analytical form of the final dynamic equations without any consequence to
our qualitative study. Recent studies show that the quartic distortion
influences the rotational properties \cite{Fetter} and the collective
excitations \cite{Li} of BECs. An interesting question is whether the
additional quartic distortion influences the structure and stability of a
DBFM.

Eqs. (6) and (7) can be renormalized to dimensionless form. By introducing
the notations $x_{0}=x/a_{0}$, $y_{0}=y/a_{0}$, $\eta =t\omega _{\bot }$, $%
\rho =m_{B}/m_{F}$, $c=g_{BB}N_{B}/(\hbar \omega _{\bot }a_{0}^{2})$, $%
f=g_{BF}N_{F}/(\hbar \omega _{\bot }a_{0}^{2})$, $k=g_{BF}N_{B}/(\hbar
\omega _{\bot }a_{0}^{2})$, $\psi _{B}=a_{0}\Psi _{B}/\sqrt{N_{B}}$ and $%
\psi _{F}=a_{0}\Psi _{F}/\sqrt{N_{F}}$, we obtained the rescaled
Euler-Lagrange equations for the mixture%
\begin{eqnarray}
i\frac{\partial }{\partial \eta }\psi _{B} &=&[-\frac{1}{2}(\frac{\partial
^{2}}{\partial x_{0}^{2}}+\frac{\partial ^{2}}{\partial y_{0}^{2}})+\frac{1}{%
2}(x_{0}^{2}+y_{0}^{2})+\frac{\lambda }{2}(x_{0}^{2}+y_{0}^{2})^{2}  \notag
\\
&&+c\left\vert \psi _{B}\right\vert ^{2}+f\left\vert \psi _{F}\right\vert
^{2}]\psi _{B},  \label{rescaled boson equation} \\
i\frac{\partial }{\partial \eta }\psi _{F} &=&[-\frac{\rho }{6}(\frac{%
\partial ^{2}}{\partial x_{0}^{2}}+\frac{\partial ^{2}}{\partial y_{0}^{2}})+%
\frac{1}{2}(x_{0}^{2}+y_{0}^{2})+\frac{\lambda }{2}(x_{0}^{2}+y_{0}^{2})^{2}
\notag \\
&&+\mu _{0F}+k\left\vert \psi _{B}\right\vert ^{2}]\psi _{F},
\label{rescaled fermion equation}
\end{eqnarray}%
where $n_{j0}=\left\vert \psi _{j}\right\vert ^{2}(j=B,F)$ are the rescaled
bosonic and fermionic densities. Here the rescaled bulk chemical potential
of the Fermi gas with $\gamma =\omega _{z}/\omega _{\bot }$ reads%
\begin{equation}
\mu _{0F}=\left\{
\begin{array}{ll}
2\pi \rho N_{F}\left\vert \psi _{F}\right\vert ^{2},\quad \quad 0\leqslant ({%
\frac{1}{\gamma })}\rho N_{F}\left\vert \psi _{F}\right\vert ^{2}<{\frac{1}{%
2\pi },} &  \\
\sqrt{4\pi \rho \gamma N_{F}\left\vert \psi _{F}\right\vert ^{2}-\gamma ^{2}}%
{,}\quad ({\frac{1}{\gamma })}\rho N_{F}\left\vert \psi _{F}\right\vert
^{2}\geqslant {\frac{1}{2\pi }}. &
\end{array}%
\right.  \label{rescaled chemical potential}
\end{equation}

For the case of a DBFM sustaining a bosonic vortex, a mixing-demixing
transition and a collapse in the mixture have been predicted by Adhikari
\textit{et al }\cite{Adhikari3}. In view of the recent studies of VAVSS in
BECs \cite{Kapale,Liu,Simula,Thanvanthri,Wen1,Andersen,Wright} it is of
interest to see how the structure and the stability of a DBFM modify when
the Bose gas is in a VAVSS. The appearance of a quantized VAVSS as well as a
pure vortex state is the genuine confirmation of phase coherence and
superfluidity of the boson component in the DBFM. As the VAVSS can stably
exist in nonrotated BECs \cite{Kapale,Liu}, we consider the bosonic VAVSS of
the dimensionless form \cite{Wen1}%
\begin{equation}
\psi _{B}=Ae^{-(x_{0}^{2}+y_{0}^{2})/2\sigma ^{2}}e^{i\delta }[\alpha
(x_{0}+iy_{0})^{l}+\beta e^{i\phi }(x_{0}-iy_{0})^{l}],
\label{bosonic VAVSS}
\end{equation}%
where $A$ is the normalization constant, $\sigma $ denotes the width of BEC,
$\delta $ is a constant phase which may be taken to be zero without loss of
generality, and $l$ is the winding number describing the quantum
circulation. The real parameters $\alpha ,\beta $ characterize the
proportion of the vortex and antivortex with $\alpha ^{2}+\beta ^{2}=1$. The
relative phase $\phi $ between the vortex and the antivortex just causes
offset of the density distribution by an angle $\phi /2l$.

\section{Results and discussion \ \ }

Here we numerically solve the coupled quantum-hydrodynamic equations (\ref%
{rescaled boson equation}) and (\ref{rescaled fermion equation}) with bulk
chemical potential $\mu _{0F}$ given by Eq. (\ref{rescaled chemical
potential}) and a trial bosonic VAVSS with definite ratio of $\alpha ^{2}$
and $\beta ^{2}$. The trial solution of the fermionic component is chosen as
a Gaussian wave function. In view of the petal-like structure of the VAVSS
in BECs, we perform exact 2D numerical calculations which require an
enormous computation effort. In our numerical simulation, we consider the $%
^{87}$Rb-$^{40}$K mixture with $m_{B}$ being the mass of $^{87}$Rb atom and $%
m_{F}$ the mass of $^{40}$K atom, and take $\omega _{z}/2\pi =100$ Hz, $%
\gamma =10$. In order to investigate the structure and stability of the DBFM
with a bosonic VAVSS, we seek the stationary solution of Eqs. (\ref{rescaled
boson equation}) and (\ref{rescaled fermion equation}) in the presence of a
bosonic VAVSS (\ref{bosonic VAVSS}). There are several typical numerical
approaches for obtaining the stationary solution of a nonlinear Sch\"{o}%
dinger equation, for instance, the Newton relaxation method \cite{Zhang} and
the imaginary time propagation (ITP) method \cite{Liu,Wen2}. In this paper,
we obtain the ground state of the DBFM with a bosonic VAVSS by using the ITP
method based on the split-step Fourier algorithm \cite{Wen3}. Our
preliminary study shows that the petal structure of a VAVSS with $\alpha
^{2}\neq \beta ^{2}$ only emerges under the condition of extremely weak
interatomic interaction or very small particle number, which is usually not
met in experiments. For a VAVSS with $\alpha ^{2}=\beta ^{2}=1/2$, however,
the petal structure always appears for arbitrary value of the interatomic
interaction or the particle number. So we shall mainly consider the case of $%
\alpha ^{2}=\beta ^{2}=1/2$. The relative phase is taken to be $\phi =0$
throughout this paper. The BB and BF interactions can be arbitrarily tuned
by the Feshbach resonance technique through varying a background magnetic
field \cite{Pethick}.

Firstly, we consider the solutions of system in the absence of interspecies
interaction. In figure 1 we show the 2D bosonic density distribution $%
n_{B0}(x_{0},y_{0})$ (top), the 2D fermionic one $n_{F0}(x_{0},y_{0})$
(middle), and the column density distributions $n_{j0}(x_{0},y_{0}=0)$ $%
(j=B,F)$ of the bosonic and fermionic components (bottom) in a DBFM with a
bosonic VAVSS, where $N_{B}=1000$, $N_{F}=100$, $l=1$, $\alpha ^{2}=\beta
^{2}=1/2$, $a_{BB}=40$ nm, $a_{BF}=0$, and the anharmonic parameters are $%
\lambda =0$ (left) and $\lambda =0.1$ (right), respectively. The bosonic
component in the mixture exhibits a petal-like structure (see figures 1(a)
and (d)), which is similar to the case of a pure BEC. At the same time the
fermionic component displays a Gaussian distribution due to the absence of
BF interaction, as shown in figures 1(b) and (e). Comparing with the case of
a harmonic potential ($\lambda =0$, see figures 1(a) and (b)), the density
distributions $n_{B0}(x_{0},y_{0})$ and $n_{F0}(x_{0},y_{0})$ in the
presence of an anharmonic term ($\lambda =0.1$, see figures 1(d) and (e)),
i.e., in the presence of a quartic distortion, become more tightly confined.
This can also be seen in the column density distributions $%
n_{j0}(x_{0},y_{0}=0)$ $(j=B,F)$ along the $x_{0}$ axis (figures 1(c) and
(f)). Thus the additional quartic distortion may favor the sympathetic
cooling between the bosons and the fermions, and makes the mixture more
stable. Our numerical simulation shows that when $\lambda \neq 0,0.1$ the
density distributions $n_{B0}(x_{0},y_{0})$ and $n_{F0}(x_{0},y_{0})$ are
similar to those in the case of $\lambda =0$, i.e., the basic structure of
the DBFM is not affected by the quartic distortion. The conclusion is the
same in the presence of BF interaction. So in the following we illustrate
the equilibrium properties of the DBFM with a bosonic VAVSS with $\lambda =0$%
.

\begin{figure}[tb]
\centerline{\includegraphics[width=7.6cm]{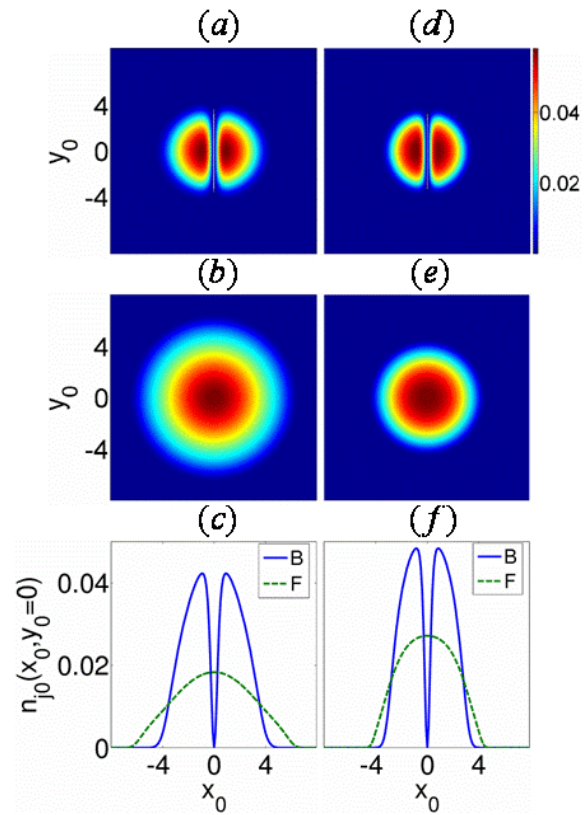}}
\caption{(color online) Bosonic density distribution $n_{B0}(x_{0},y_{0})$
(top), fermionic density distribution $n_{F0}(x_{0},y_{0})$ (middle), and
column density distributions $n_{j0}(x_{0},y_{0}=0)$ $(j=B,F)$ (bottom) for
a DBFM with a bosonic VAVSS in the absence of BF interaction. The anharmonic
parameters are $\protect\lambda =0$ (left) and $\protect\lambda =0.1$
(right), respectively. Here $N_{B}=1000$, $N_{F}=100$, $l=1$, $\protect%
\alpha ^{2}=\protect\beta ^{2}=1/2$, and $a_{BB}=40$ nm.}
\end{figure}

When the BF interaction is nonzero, the DBFM with a bosonic VAVSS can
display rich phase structures. Figure 2 gives the density distribution $%
n_{F0}(x_{0},y_{0})$ of fermions (left) and the column density distributions
$n_{j0}(x_{0},y_{0}=0)$ $(j=B,F)$ of bosons and fermions (right) for various
BF interactions, where $\lambda =0,$ $N_{B}=1000$, $N_{F}=100$, $l=1$, $%
\alpha ^{2}=\beta ^{2}=1/2$, $a_{BB}=40$ nm. As shown in figures 2(a) and
(d), for repulsive $a_{BF\text{ }}=90$ nm the system is in a partial mixed
state, where the bosonic component with a petal-like structure lies inside
the fermionic one while the fermionic cloud shows a honeycomb-like structure
due to the BF repulsion. For a sufficiently large repulsive $a_{BF\text{ }%
}=190$ nm, the fermionic cloud is completely expelled outside the BEC (see
figures 2(b) and (e)), which is usually referred as demixing, i.e., the
system is in a complete phase-separated state. The phenomenon of
mixing-demixing transition has also been found in previous theoretical
studies of a DBFM \cite{Molmer,Capuzzi,Takeuchi} or a DBFM with a bosonic
vortex \cite{Adhikari3}. For the DBFM with a bosonic VAVSS, however, with
the increase of BF repulsion the DFG can not be completely expelled from the
central region of the harmonic trap due to the bosonic VAVSS. In the case of
attractive BF interaction ($a_{BF}<0$) we find that the fermionic component
is pulled inside the bosonic one and the maximum mixing between the two
components is then achieved for a critical value of $a_{BF}$. With the
further increase of the strength of the attractive BF interaction the system
collapses. In figures 2(c) and (f), where $a_{BF\text{ }}=-90$ nm, just
below the threshold for collapse, we can see an almost complete mixing
between the bosonic and fermionic clouds. In this mixed state, the fermionic
density distribution developes a petal-like structure reminiscent of a VAVSS
as in the bosonic component. However, near the center of the trap the
fermionic density $n_{F0}(x_{0}=0,y_{0})$ tends to a nonzero constant value
and does not have the VAVSS behavior. For a VAVSS with $l=1$ and $\alpha
^{2}=\beta ^{2}=1/2$, the probability density in the gap regions between two
petals should be zero. Therefore the fermionic component is not really in a
VAVSS but tends to mimic a bosonic VAVSS due to mixing.

\begin{figure}[tb]
\centerline{\includegraphics[width=7.5cm]{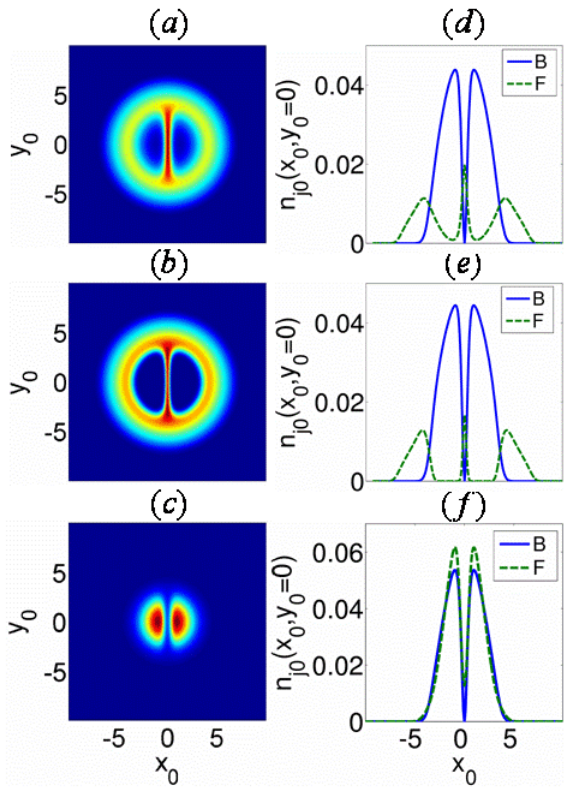}}
\caption{(color online) Fermionic density distribution $n_{F0}(x_{0},y_{0})$
(left) and column density distributions $n_{j0}(x_{0},y_{0}=0)$ $(j=B,F)$
(right) of a DBFM in the presence of a bosonic VAVSS, where $\protect\lambda %
=0$, $N_{B}=1000$, $N_{F}=100$, $l=1$, $\protect\alpha ^{2}=\protect\beta %
^{2}=1/2$, and $a_{BB}=40$ nm. The BF scattering lengths are $a_{BF}=90$ nm
(top), $a_{BF}=190$ nm (middle), and $a_{BF}=-90$ nm (bottom), respectively.}
\end{figure}

Depending on the choice of parameters the separated phase may exhibit
different configurations. Shown in figure 3 are the bosonic density
distribution $n_{B0}(x_{0},y_{0})$ (top), the fermionic one$\
n_{F0}(x_{0},y_{0})$ (middle), and the column density distributions $%
n_{j0}(x_{0},y_{0}=0)$ $(j=B,F)$ (bottom) in a DBFM with a bosonic VAVSS for
the case of equal particle numbers $N_{B}=N_{F}=1000$. Here $\lambda =0,$ $%
l=1$, $\alpha ^{2}=\beta ^{2}=1/2$, $a_{BB}=2000$ nm, and the BF $s$-wave
scattering lengths are $a_{BF}=500$ nm (left) and $a_{BF}=2000$ nm (right),
respectively. It is easy to confirm that the diluteness condition of the
gases is satisfied for all the parameters mentioned in this paper. From
figures 3(a)-(c), we can see for $a_{BF}=500$ nm and given parameters the
system is in a perfect mixed state, where the fermionic density reaches the
maximum in the gap region between two bosonic petals because of the BF
repulsion. For sufficiently strong BF repulsion $a_{BF}=2000$ nm, the bosons
are completely expelled from the center of the trap as shown in figure 3(d),
forming a ring-shaped joint \textquotedblleft shell\textquotedblright\
around the fermions due to the presence of the VAVSS. At the same time the
fermionic component develops an onion-like core, which can be seen in figure
3(e). The column density distributions $n_{j0}(x_{0},y_{0}=0)$ $(j=B,F)$ are
illustrated in figure 3(f). Figures 3(d)-(f) definitely indicate a new
separated phase of the DBFM which is different from that in figures 2(b) and
(e), where though the fermions are completely expelled outside the bosons
there is remarkable fermionic density along the axis\ of $x_{0}=0$ in the
trap center.

\begin{figure}[tb]
\centerline{\includegraphics[width=7.4cm]{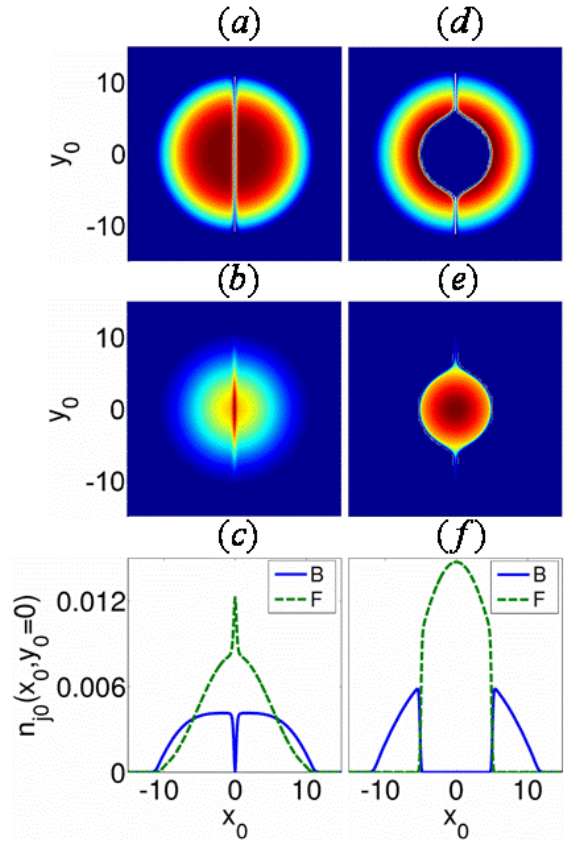}}
\caption{(color online) Bosonic density distribution $n_{B0}(x_{0},y_{0})$
(top), fermionic density distribution $n_{F0}(x_{0},y_{0})$ (middle), and
column density distributions $n_{j0}(x_{0},y_{0}=0)$ $(j=B,F)$ (bottom) for
a DBFM with a bosonic VAVSS in the case of $N_{B}=N_{F}=1000$. The BF $s$%
-wave scattering lengths are $a_{BF}=500$ nm (left) and $a_{BF}=2000$ nm
(right), respectively. The other parameters are $l=1$, $\protect\alpha ^{2}=%
\protect\beta ^{2}=1/2$, and $a_{BB}=2000$ nm.}
\end{figure}

In figure 4, we display the bosonic and fermionic density distributions $%
n_{j0}(x_{0},y_{0})$ $(j=B,F)$ as well as the column density distributions $%
n_{j0}(x_{0},y_{0}=0)$ $(j=B,F)$ for the case of $N_{B}=100\ $and $%
N_{F}=1000 $, where the interspecies $s$-wave scattering lengths are $%
a_{BF}=800$ nm (left) and $a_{BF}=8000$ nm (right), respectively. The other
parameters are $\lambda =0,$ $l=1$, $\alpha ^{2}=\beta ^{2}=1/2$,\ and $%
a_{BB}=800$ nm. For repulsive $a_{BF\text{ }}=800$ nm the DBFM begins to
evolve into a coexistence phase of component mixing and separation, where
the bosonic component forms two distant and segregated \textquotedblleft
islands\textquotedblright\ embedded in the disc-shaped fermionic gas (see
figures 4(a)-(c)). When $a_{BF}=8000$ nm the two bosonic \textquotedblleft
islands\textquotedblright\ become more compact and closer, and the fermionic
component is completely separated from the bosonic one (see figures
4(d)-(f)). As shown in figure 4(f), the fermionic component constitutes both
a \textquotedblleft shell\textquotedblright\ around and a \textquotedblleft
core\textquotedblright\ inside the BEC along the $x_{0}$-direction, i.e.,
the Fermi gas distributes on both sides of the Bose gas along the $x_{0}$%
-direction, which is similar to that found in Ref. \cite{Molmer}. However,
the layer distribution of the present separated phase is only along the $%
x_{0}$-direction in the nearby region of $y_{0}=0$ and is absent along the $%
y_{0}$-direction, i.e., the present separated phase does not show 2D or 3D
layer configuration. Furthermore, the core structure or the core plus shell
structure of the fermionic cloud in a DBFM without vortex can be achieved by
varying the BF interaction continuously \cite{Molmer}, whereas the different
separated phases in the DBFM with a bosonic VAVSS correspond respectively
the different parameter conditions and can not be achieved by merely change
the BF coupling continuously. The above analysis indicates that the inlay
configuration of the two bosonic \textquotedblleft
islands\textquotedblright\ embedded in the disc-shaped Fermi gas is a novel
separated phase. We expect that this novel separated phase can be observed
and tested in the future experiments. For large $a_{BF}$ the fermions could
possibly form $p$-wave pairing, which is beyond the scope of the mean-field
treatment. The two compact bosonic segregated \textquotedblleft
islands\textquotedblright\ in figure 4(d) perhaps can be understood as a
consequence of buoyancy in the Fermi sea.

\begin{figure}[tb]
\centerline{\includegraphics[width=7.4cm]{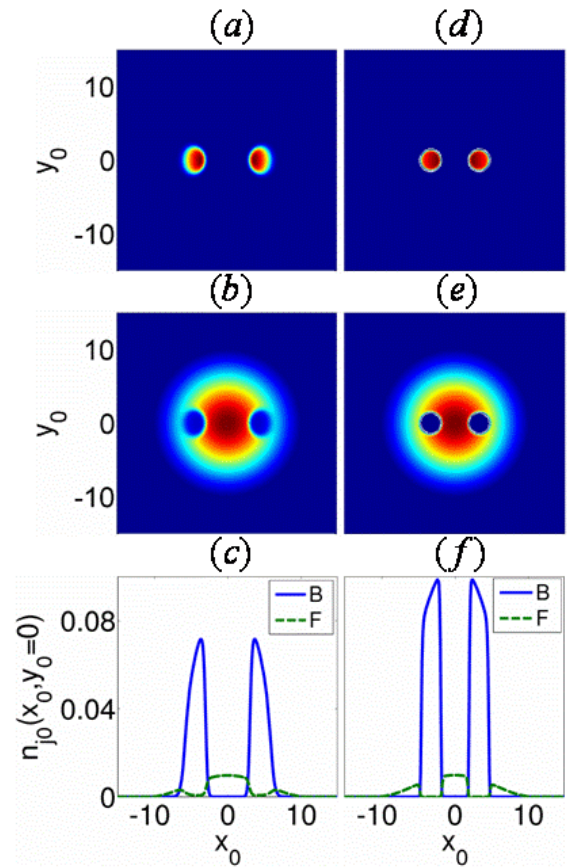}}
\caption{(color online) Bosonic density distribution $n_{B0}(x_{0},y_{0})$
(top), fermionic density distribution $n_{F0}(x_{0},y_{0})$ (middle), and
column density distributions $n_{j0}(x_{0},y_{0}=0)$ $(j=B,F)$ (bottom) for
a DBFM with a bosonic VAVSS in the case of $N_{B}=100$ and $N_{F}=1000$. The
BF $s$-wave scattering lengths are $a_{BF}=800$ nm (left) and $a_{BF}=8000$
nm (right), respectively. The other parameters are $l=1$, $\protect\alpha %
^{2}=\protect\beta ^{2}=1/2$, and $a_{BB}=800$ nm.}
\end{figure}

The BF interaction has a very strong influence not only on the structure of
the DBFM with a bosonic VAVSS but also on the stability of the system. For a
repulsive BF interaction (positive $a_{BF}$), a stable structure can always
be achieved for a fixed $N_{B}$, $N_{F}$, $\lambda $, $l$, $\alpha $, $\beta
$, $\phi $, and BB scattering length $a_{BB}$ as shown in figures 2-4.
However, for a sufficiently strong attractive BF interaction (negative $%
a_{BF}$), the system can undergo a simultaneous collapse of the bosonic and
fermionic density distributions. Physically, the critical value of BF
scattering length is determined by the balance between the repulsion of
bosons and fermions and the mutual attractive BF interaction. Likewise, when
the boson number $N_{B}$ or fermion number $N_{F}$ are sufficiently large,
the attractive BF interaction can not be stabilized by the repulsions in BB
and BF subsystems. Thus the mixture lowers its energy via increasing the
boson and fermion densities, and finally the bosonic component or the
fermionic one or both the bosonic and fermionic components collapse
simultaneously due to instability. In our investigation, the instability
signature is found by monitoring the failure of the numerical iterative
process, i.e., the invalidation of the coupled equations (\ref{rescaled
boson equation}) and (\ref{rescaled fermion equation}) describing the DBFM
with conserved quantities $N_{B}$ and $N_{F}$. In fact, an indefinite growth
of the maximum of the bosonic and fermionic densities always indicates the
appearance of an instability onset, and therefore the collapse of the
mixture is triggered and unavoidable. This approach is accurate and reliable
\cite{Jezek}.

The stability region in the plane spanned by $\log (N_{F})$ and $\log
(N_{B}) $ is shown in figure 5. As an illustration, we assume the BB and BF
scattering lengths of the $^{87}$Rb-$^{40}$K mixture are $a_{BB}=5$ nm and $%
a_{BF}=-15$ nm which are very close to the typical experimental values $%
a_{BB}=5.25$ nm and $a_{BF}=-13.8$ nm reported by Ferrari \textit{et al }%
\cite{Ferrari}. The four different curves (square, circle, triangle, and
inverted triangle) in the figure 5 mark the stability limit for four
different cases of DBFM: (i) DBFM without vortex, (ii) DBFM with a bosonic
VAVSS of $l=1$ and $\alpha ^{2}=\beta ^{2}=1/2$, (iii) DBFM with a bosonic
VAVSS of $l=1$, $\alpha ^{2}=3/4$ and $\beta ^{2}=1/4$, and (iv) DBFM with a
bosonic vortex of $l=1$, respectively. The region below each curve denotes
the stability region of the mixture, and that above each curve corresponds
to the instability (collapse) one of the system. The stability curves in the
range of $\log (N_{F})\leqslant 2$ are four almost overlapped and parallel
horizontal lines (data not shown here) which are similar to those in the
range of $\log (N_{F})\geqslant 3.9$.

From figure 5 we can see for the same parameters the DBFM with a bosonic
vortex is more stable than the one without vortex due to the repulsive
centrifugal kinetic energy, in agreement with previous study \cite{Adhikari3}%
. In addition, for the case of $N_{F}\leqslant $ int$(10^{3.8})$ $\approx
6310$ the DBFM with a bosonic vortex is more stable than that with a bosonic
VAVSS because the energy of a VAVSS (especially, a VAVSS with equal
proportions $\left\vert \alpha \right\vert =\left\vert \beta \right\vert $)
is higher than that of a pure vortex (or antivortex) for a given $l$ \cite%
{Liu}. However, for the case of $6310<N_{F}<$ int$(10^{3.9})\approx 7943$ we
find that the DBFM with a bosonic VAVSS is more stable than that with a
bosonic vortex. Notice that for attractive BF interaction the fermionic
density distribution tends to simulate the bosonic one (see figure 2(c)),
thus the overlap region between the bosons and fermions in a DBFM with a
bosonic VAVSS is smaller than that in a DBFM with a bosonic vortex due to
the petal-like structure of VAVSS. This effect dominates the stability of
the system when the number of fermions is sufficiently large such that the
influence of the BF attraction on the collapse of a DBFM with a bosonic
VAVSS is weaker than that of a DBFM with a bosonic vortex, which is a
probable reason for the above intermittency phenomenon in the stability
region. At the same time, there is also an intermittency phenomenon between
the stability curve for the DBFM with a bosonic VAVSS of $\alpha ^{2}=1/2$
and that of $\alpha ^{2}\neq 1/2$ (e.g., $\alpha ^{2}=3/4$). The analysis is
similar to the above discussion. Note that the stability curve for the DBFM
with a bosonic VAVSS of $l=1$, $\alpha ^{2}=1/4$ and $\beta ^{2}=3/4$ is the
same with that of $l=1$, $\alpha ^{2}=3/4$ and $\beta ^{2}=1/4$ because the
two systems have the equal energy \cite{Liu}, which is verified in our
numerical simulation. Furthermore, the DBFM with a bosonic VAVSS ($l=1$) is
more stable than the conventional one without vortex ($l=0$) in the wide
range of $N_{F}<7943$ because for the bosonic VAVSS there is an additional
repulsive centrifugal term $\hbar ^{2}l^{2}/(2m_{B}(x^{2}+y^{2}))$ in the
stationary GP equation \cite{Liu}. When $N_{F}\geqslant 7943$ all the four
different cases of DBFM become instable for $N_{B}$ above the order of unit.
For the DBFMs with a bosonic vortex (VAVSS) of higher $l$, our simulation
shows that the stability curves are similar to the case of $l=1$, but the
systems of higher $l$ are less stable than those of $l=1$.

\begin{figure}[tb]
\centerline{\includegraphics[width=7.6cm]{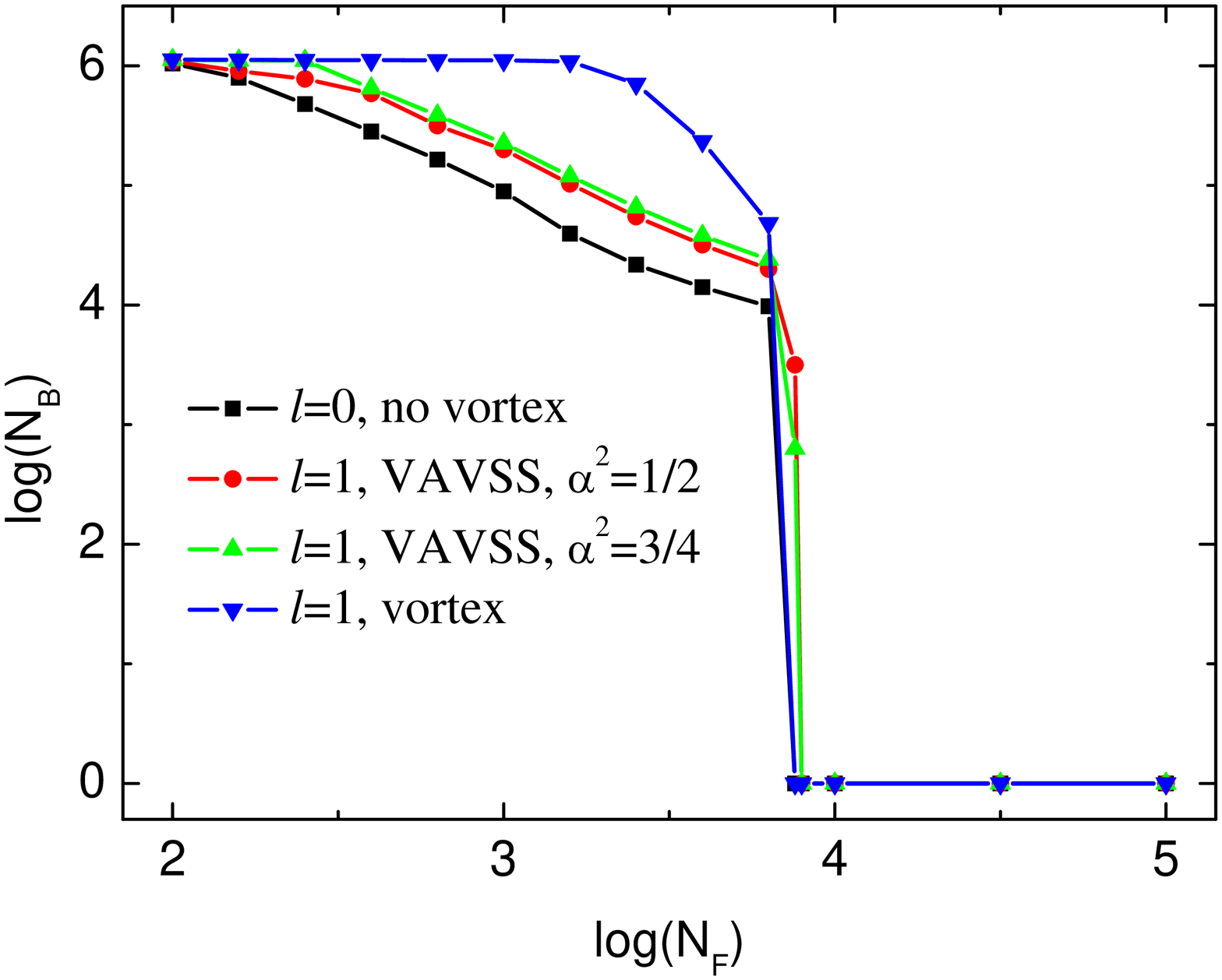}}
\caption{(color online) Stability region for the degenerate $^{87}$Rb-$^{40}$%
K mixture with scattering lengths $a_{BB}=5$ nm and $a_{BF}=-15$ nm. Here
the curves of square, circle, triangle, and inverted triangle show the
stability limit for the DBFM without vortex, the DBFM with a bosonic VAVSS
of $l=1$ and $\protect\alpha ^{2}=\protect\beta ^{2}=1/2$, the DBFM with a
bosonic VAVSS of $l=1$, $\protect\alpha ^{2}=3/4$ and $\protect\beta %
^{2}=1/4 $, and the DBFM with a bosonic vortex of $l=1$, respectively. For $%
\log (N_{B})$ above the limit or $\log (N_{F})$ on the right of the limit
the bosonic component or the fermionic one or both the two components
collapse simultaneously. }
\end{figure}

\section{Conclusion}

We have investigated the structure and stability of a quasi-2D DBFM with a
bosonic VAVSS, where the BB interaction is taken to be repulsive and the BF
interaction to be both repulsive and attractive. It is shown that the
bosonic and fermionic density distributions for a DBFM with a bosonic VAVSS
in a harmonic plus quartic potential are more tightly confined than those in
a harmonic potential. The additional quartic distortion due to experimental
uncertainties or artificial quartic confinement does not influence the
stationary structure of the mixture, and may favor the sympathetic cooling
between the interspecies and makes the system more stable.

Depending on the choice of parameters, the DBFM with a bosonic VAVSS can
show rich phase structures. For repulsive BF interaction, the bosonic
component can form a petal-shaped \textquotedblleft core\textquotedblright\
inside the honeycomb-like fermionic one, or a ring-shaped joint
\textquotedblleft shell\textquotedblright\ around the onion-like fermionic
cloud, or multiple segregated \textquotedblleft islands\textquotedblright\
embedded in the disc-shaped Fermi gas. Note that the different separated
phases in the DBFM with a bosonic VAVSS are formed under different parameter
conditions and can not be achieved by just changing the BF coupling
continuously, which is different from the case of a conventional DBFM \cite%
{Molmer}. In addition, there is a mixing-demixing transition which is
controlled by the strength of the BF repulsion. For attractive BF
interaction below the threshold for collapse, an almost complete mixing
between the bosonic and fermionic components can be formed, where the
fermionic component tends to simulate a bosonic VAVSS. Furthermore, we give
the stability region for four different cases of DBFM with specific
parameters, where the two stability curves of the DBFM with a bosonic VAVSS
and the one with a bosonic vortex display an intermittent configuration.
These new findings can be verified in the future experiments on DBFMs, and
thus the present work provides a new way to test further the validity and
prediction of the quantum-hydrodynamic model.

\section{Acknowledgments}

The authors sincerely thank Prof. Biao Wu for valuable comments and
discussions. This work was supported by the NSFC under Grant No. 10847143,
the NSF of Shandong Province under Grant No. Q2007A01, and PhD Foundation of
Liaocheng University.

\end{document}